\documentclass[12pt,numreferences]{kluwer}

\usepackage{dsfont}
\usepackage{epsfig}

\begin{document}

\begin{opening}
\title{Dispersive calculation of the\\
massless multi-loop
sunrise diagram}
\author{ANDREAS \surname{ASTE}}
\institute{Department of Physics and Astronomy, Theory Division,
University of Basel, Klingelbergstrasse 82, 4056 Basel, Switzerland\\
e-mail: andreas.aste@unibas.ch}
\runningtitle{MASSLESS SUNRISE DIAGRAMS}
\runningauthor{ANDREAS ASTE}

\begin{abstract}
The massless sunrise diagram with an arbitrary number of
loops is calculated in a simple but formal manner.
The result is then verified by rigorous mathematical treatment.
Pitfalls in the calculation with distributions are highlighted and
explained. The result displays the high energy behaviour of the massive
sunrise diagrams, whose calculation is involved already for the two-loop
case.
\end{abstract}
\classification{AMS classification (2000)}{81T05,81T15,81T18.}

\end{opening}

\section{Introduction}
The Feynman propagator of a free massless scalar quantum field
$\Phi(x)$ fulfilling the wave equation
\begin{equation}
\Box \Phi(x) = \partial_\mu \partial^\mu \Phi(x)=0
\end{equation}
is given in configuration space by the vacuum expectation value
\begin{displaymath}
\Delta_F(x)=-i \langle 0 | T (\Phi(x) \Phi(0)) | 0 \rangle
=\int \frac{d^4 k}{(2 \pi)^4} \frac{e^{-ikx}}{k^2+i0}
\end{displaymath}
\begin{equation}
=\frac{i}{4 \pi^2} \frac{1}{x^2-i0} = \frac{i}{4 \pi^2} P \frac{1}{x^2}-
\frac{1}{4 \pi} \delta (x^2),
\end{equation}
where $T$ is the time-ordering operator, $P$ denotes principal value
regularization and $\delta$ is the one-dimensional Dirac distribution depending
on $x^2=x_\mu x^\mu=(x^0)^2-(x^1)^2-(x^2)^2-(x^3)^2=x_0^2-{\vec{x}}^2$.
Formally, one could write for the massless $(n-1)$-loop sunrise diagram
\begin{displaymath}
\int d^4 x [ \Delta_F(x)]^n e^{ikx} = \Gamma_n(k)=
\end{displaymath}
\begin{equation}
\int
\frac{d^4 k_1}{(2 \pi)^4} \ldots \frac{d^4 k_{n-1}}{(2 \pi)^4}
\frac{1}{k_1^2+i0}  \frac{1}{k_2^2+i0} \ldots \frac{1}{(k-k_1-k_2-\ldots -
k_{n-1})^2+i0}. \label{sunrise}
\end{equation}

\begin{figure}[htb]
\begin{center}
\includegraphics[height=4.2cm]{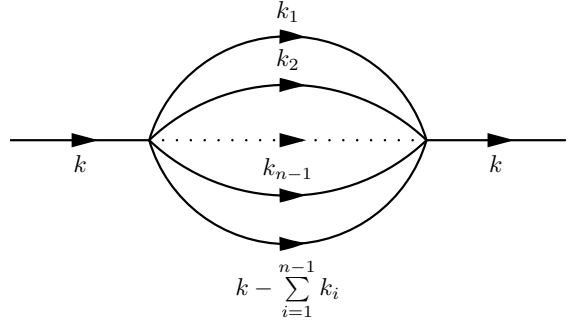}
\caption{The (n-1)-loop sunrise diagram.}
\end{center}
\end{figure}

This expression above is both ultraviolet and infrared divergent for $n \ge 2$, consequently
the corresponding expressions in eq. (\ref{sunrise}) are all ill-defined.
Therefore, $[ \Delta_F(x)]^n$ has to be regularized in order to obtain
a mathematically well-defined tempered distribution in the dual Schwartz space
${\bf{\mathcal{S}}'}(\mathds{R}^4)$. The regularization
can be performed by one of the many well-known methods
\cite{Pauli,Dim,Schwinger,GScharf,eg,Aste},
which all lead to the same result. For $n=2$ one obtains
\begin{displaymath}
\Gamma_2^{\mathcal{R}}(k)= \mathcal{R} \Biggl[
\int \frac{d^4 k_1}{(2 \pi)^4} \frac{1}{k_1^2+i0} \, \frac{1}{(k-k_1)^2+i0}
\Biggr]
\end{displaymath}
\begin{equation}
=-\frac{i}{4(2 \pi)^2} \log \Biggl (-\frac{k^2+i0}{\mu_{\mathcal{R}}^2} \Biggr)=
-\frac{i}{4(2 \pi)^2} \log|k^2/\mu_{\mathcal{R}}^2|-
\frac{1}{16 \pi} \Theta(k^2) \, , \label{gamma2}
\end{equation}
where $\mathcal{R}$ denotes the regularization procedure with renormalization
scale $\mu_{\mathcal{R}}$ and $\Theta$ is
the Heaviside distribution. Obviously, the regularized expression
for the one-loop diagram $\Gamma_2^{\mathcal{R}}(k)$ is defined up to a constant
im momentum space or up to a local distribution in configuration space $\sim \delta^{(4)}(x)$.
The scaling symmetry of the Feynman propagator $\Delta_F(\lambda x)=\lambda^{-2}
\Delta_F(x)$, $\lambda \! \in \! \mathds{R}$,
is {\em{spontaneously broken}} by the regularization procedure, such that a corresponding
scaling law $\Gamma_2^{\mathcal{R}}(\lambda k) = \Gamma_2^{\mathcal{R}}(k)$
does {\it not} hold as one might expect naively from the formal definition
of $\Gamma_2(k)$ in eq. (\ref{sunrise}).

\section{Formal approach}
In order to obtain the analytic result for an arbitrary multi-loop diagram,
a {\em{purely formal}} calculation is presented here first.
One easily derives formally for $n \neq 1$
\begin{equation}
\Box \frac{1}{(x^2-i0)^n} = \frac{4n(n-1)}{(x^2-i0)^{n+1}},
\end{equation}
and therefore
\begin{equation}
[\Delta_F(x)]^n=-\frac{i^n}{4^{n-2} (2 \pi)^{2(n-2)}
(n-1)!(n-2)!} \Box^{n-2} [\Delta_F(x)]^2 \, . \label{formal}
\end{equation}
Since the d'Alembert operator $\Box$ goes over into $-k^2$ in momentum space,
one obtains
\begin{equation}
\Gamma_n^{\mathcal{R}}(k)=-\frac{(-i)^n (k^2)^{n-2}}{4^{n-2} (2 \pi)^{2(n-2)}
(n-1)!(n-2)!} \Gamma_2^{\mathcal{R}} (k) +
\sum \limits_{i=0}^{n-2} c_i^{\mathcal{R}}
(k^2)^i \, , \quad n \ge 2, \label{result}
\end{equation}
if the assumption is made that the formal result eq. (\ref{formal})
is also valid for the regularized expressions.
The additional terms involving the Fourier transform of
(covariant derivatives of) the $\delta$-distribution in configuration space
reflect the fact that the regularized Lorentz invariant product of
Feynman propagators is only defined up to local terms given by
derivatives of the $\delta$-distribution, whose order can be
restricted by the requirement that the scaling behaviour of (\ref{result})
corresponds to the artificial degree of divergence $\omega=2(n-2)$
of the $(n-1)$-loop sunrise diagram. 
Further restrictions follow from physical
symmetry requirements, depending on the underlying theory in which the
diagram plays a role (like in perturbative quantum gravity).
As an example, for the 2-loop diagram $(n=3)$ one has the general result
\begin{equation}
\Gamma_3^{\mathcal{R}}(k)=-\frac{k^2}{32(2 \pi)^4} \log \Biggl(
-\frac{k^2+i0}{\mu_\mathcal{R}^2} \Biggr) + c_0^{\mathcal{R}}.
\end{equation}
The simple result given by eq. (\ref{result})
is indeed correct and similar formal calculations as presented above
can be found in many textbooks (see, e.g., \cite{Cheng}).
The massive two-loop case of the sunrise diagram which is already
involved from the computational point of view has been studied in detail
in numerous papers (see e.g. \cite{Kallen,TkaChet,Bauberger,Berends,
Fleischer,PostTausk,DavySmi,Laporta,Caffo2,Passarino,AsteSunrise} and
references therein), and to a certain extent also for the massive 
multi-loop case \cite{Laporta,Laporta2,Groote,GKP}.

\section{Properties of tempered distributions}
A rigorous derivation of (\ref{result}) is presented below.
For the forthcoming discussion, it is advantageous to review some of the useful properties
of basic tempered distributions appearing in quantum field theory.
A scalar neutral field $\Phi$ can be decomposed into a negative and positive frequency
part according to
\begin{equation}
\Phi(x)=\Phi^-(x)+\Phi^+(x)=
\int \frac{d^3k}{\sqrt{(2 \pi)^3 2 |\vec{k}|}} \Bigl[ a(\vec{k})
e^{-ikx}+a^\dagger(\vec{k}) e^{ikx} \Bigr] .
\end{equation}
The commutation relations for such a field are given by the
positive and negative frequency Jordan-Pauli distributions
\begin{equation}
\Delta^\pm(x)=-i [\Phi^{\mp}(x) , \, \Phi^{\pm}(0)] =
-i \langle 0 |  [\Phi^{\mp}(x) , \, \Phi^{\pm}(0)] | 0 \rangle \, ,
\label{commutator}
\end{equation}
which have the Fourier transforms
\begin{equation}
\hat{\Delta}^{\pm}(k)= \int d^4x \, \Delta^{\pm}(x) e^{ikx}=
\mp (2 \pi i) \,  \Theta(\pm k^0) \delta(k^2) . \label{dfourier}
\end{equation}
The fact that the commutator
\begin{equation}
[\Phi(x),\Phi(0)]=i\Delta^+(x)+i\Delta^-(x) =: i \Delta(x)
\end{equation}
vanishes for spacelike arguments $x^2 < 0$
due to the requirement of microcausality, leads to
the important property that the Jordan-Pauli distribution
$\Delta$ has {\emph{causal support}}, i.e. it vanishes outside the closed
forward and backward lightcone such that
\begin{equation}
\mbox{supp} \, \Delta(x) \subseteq \overline{V}^- \cup \overline{V}^+  \, , \quad
\overline{V}^\pm=\{x \, | \, x^2 \ge 0, \, \pm x^0 \ge 0 \}
\end{equation}
in the sense of distributions.
A further crucial observation is the fact that one can introduce the
retarded propagator $\Delta^{ret}(x)$ which coincides with $\Delta(x)$ on
$\overline{V}^+ \! - \{0 \}$, i.e. $\langle \Delta^{ret}, \varphi \rangle =
\langle \Delta, \varphi \rangle$ holds for all test functions in the
Schwartz space $\varphi \! \in
\! \mathcal{S}(\mathds{R}^4)$ with support $\mbox{supp} \, \varphi \subset
\mathds{R}^4 - \overline{V}^-$.
One might be tempted to write down in configuration space
\begin{equation}
\Delta^{ret}(x)=\Theta(x^0) \Delta(x) ,
\end{equation}
and to transform this expression into a convolution in momentum space
\begin{equation}
\hat{\Delta}^{ret}(k)=\int \frac{d^4 p}{(2 \pi)^4} \, \hat{\Delta}(p) \hat{\Theta}
(k-p) \, , \quad
\quad \hat{\Delta}(p)=-(2 \pi i) \mbox{sgn}(p^0) \delta(p^2) \, . \label{convo}
\end{equation}
The Heaviside distribution $\Theta(x^0)$ could be replaced by
$\Theta(vx)$ with an arbitrary
vector $v \in V^+=\overline{V}^+ \! - \partial \overline{V}^+$
inside the open forward lightcone.
The Fourier transform of the Heaviside distribution $\Theta(x^0)$
can be calculated easily
\begin{equation}
\hat{\Theta}(k)=\lim_{\epsilon \rightarrow 0}
\int d^4x \, \Theta(x^0) e^{-\epsilon x^0} e^{ik_0x^0-i \vec{k}\vec{x}}=
\frac{(2 \pi)^3 i}{k^0+i0} \delta^{(3)}(\vec{k}) .
\end{equation} 
For the special case where $k$ is in the forward lightcone $V^+$,
one can go to a Lorentz frame where $k=(k^0,\vec{0})$
such that eq.
(\ref{convo}) becomes
\begin{equation}
\hat{\Delta}^{ret}(k^0,\vec{0})=\frac{i}{2 \pi} \int dp^0 \frac{\hat{\Delta}
(p^0)}{k^0-p^0+i0}=
\frac{i}{2 \pi} \int dt \, \frac{\hat{\Delta}
(tk^0)}{1-t+i0} .
\end{equation}
Hence, for arbitrary $k \in V^+$, $\hat{\Delta}^{ret}$ would be
given by the {\emph{dispersion relation}}
\begin{equation}
\hat{\Delta}^{ret}(k)=\frac{i}{2 \pi} \int dt \, \frac{\hat{\Delta}
(tk)}{1-t+i0} . \label{disprel}
\end{equation}
However, the integral in eq. (\ref{disprel}) is undefined. One can circumvent this
problem, e.g., by introducing a mass for the field $\Phi$ as a regulator for the
$\Delta$-distribution. For the massive Jordan-Pauli distribution $\Delta_m(x)$
one gets due to the $\delta$-distribution in
\begin{equation}
\hat{\Delta}_m(k)=-(2 \pi i) \mbox{sgn}(k^0) \delta(k^2-m^2)
\end{equation}
the expression ($k \! \in \! V^+$)
\begin{displaymath}
\hat{\Delta}_m^{ret}(k)=\int dt \, \frac{
\mbox{sgn}(tk^0) \delta(t^2k^2-m^2)}{1-t+i0}=
\end{displaymath}
\begin{equation}
\int dt \, \frac{
\bigl[ \delta(t-\frac{m}{\sqrt{k^2}})- \delta(t+\frac{m}{\sqrt{k^2}})
\bigr]}{2 \sqrt{k^2} m (1-t+i0) }=
\frac{1}{k^2-m^2} \quad (k^2 \ne m^2) \, .
\end{equation}
As a special case of the edge of the wedge theorem \cite{PCT}
it is known that the Fourier transform of the retarded distribution
$\hat{\Delta}^{ret}(k)$ is the boundary value of an analytic function
$r(z)$, regular in $T^+ := \mathds{R}^4+ i V^+$. This way one obtains
from $r(z)=1/(z^2-m^2)$, $z \! \in \! T^+$, and $m \rightarrow 0$
\begin{equation}
\hat{\Delta}^{ret}(k)=\frac{1}{k^2+ik^0 0}.
\end{equation}
The analytic expression for the Feynman propagator is then recovered,
which coincides with
\begin{equation}
\hat{\Delta}^{ret}(k)=\hat{\Delta}_F(k)+\Delta^-(k)=\frac{1}{k^2+ik^0 0}
\end{equation}
for $k \in V^+$.
The singular part of the Feynman propagator is given by
\begin{equation}
\mbox{Im} \, (\hat{\Delta}_F(k))=- \pi \delta(k^2)
\end{equation}
and can be deduced directly from the causal
Jordan-Pauli distribution $\hat{\Delta}$ in an obvious way.

\section{Analysis of causal properties}
The causal properties of the Jordan-Pauli distribution and the retarded
propagator can be observed in an analogous way in the case
of the distributions describing a multi-loop sunrise diagram.
The normally ordered product of $n$ free field operators $:\Phi(x)^n:$
is an operator valued distribution.
Therefore, the same is true for the tensor product 
$:\Phi(x)^n:$$:\Phi(y)^n:$ \cite{Constantinescu}.
The vacuum expectation value
of ${\langle 0| \! :\Phi(x)^n::\Phi(0)^n: \! | 0 \rangle}$ can be extracted by Wick's
theorem and leads to $i^n n! \Delta^+(x)^n$ (the
combinatorial factor $n!$ will be disregarded in the following),
which is indeed a well-defined tempered distribution.
Calculating the retarded part of the causal distribution $\Delta_n(x)$ which we define
for $ n \ge 2$ by
\begin{equation}
\Delta_n(x):=i^n(\Delta^+(x)^n-(-1)^n(\Delta^-(x)^n)=
\frac{1}{n!} \langle 0 | [:\Phi(x)^n:,:\Phi(0)^n:] | 0 \rangle \, , \label{dn}
\end{equation}
which has causal support, will finally lead to the desired result eq. (\ref{result}).
For this purpose, the Fourier transform of $\Delta_n(x)$ has to be calculated.
This can be done inductively. For $n=2$, we first evaluate the Fourier
transform of $\Delta_2^-(x):=\Delta^-(x)^2$ given by the convolution
\begin{displaymath}
{\hat \Delta}_2^-(k)= \int \frac{d^4 q}{(2 \pi)^4} {\hat \Delta}^- (q) {\hat \Delta}^- (k-q)=
\end{displaymath} 
\begin{equation}
-\frac{1}{(2 \pi)^2} \int d^4 q \, \Theta(-q^0) \delta(q^2)
\Theta(q^0-k^0) \delta((k-q)^2). \label{I2}
\end{equation}
The integral vanishes outside the closed backward lightcone
due to Lorentz invariance and the two $\Theta$-distributions in eq. (\ref{I2}).
Therefore one can go to a Lorentz frame where $k=(k^0<0, \vec{0})$,
and using the abbreviation $E=\sqrt{\vec{q}^{\, 2}}=|\vec{q} \, |$ and exploiting the
$\delta$-distribution in eq. (\ref{I2}) one obtains
\begin{displaymath}
{\hat \Delta}_2^-(k^0 \! < \! 0,\vec{0})=-\frac{1}{(2 \pi)^2} \int \frac{d^3 q}{2E}
\delta((k^0)^2+2 k^0 E)\Theta(-E-k^0)=
\end{displaymath}
\begin{equation}
-\frac{1}{2(2 \pi) k^0} \int \! d |\vec{q} \, | |\vec{q} \, |
\delta\Bigl(\frac{k^0}{2}+|\vec{q} \, |\Bigr) \Theta(-|\vec{q} \, |-k^0)=
\! -\frac{1}{4(2 \pi)} \Theta(-k^0) \Theta((k^0)^2),
\end{equation}
and for arbitrary $k$ one gets the intermediate result
\begin{equation}
{\hat \Delta}_2^-(k)=-\frac{1}{8 \pi} \Theta(-k^0) \Theta(k^2).
\end{equation}
According to eq. (\ref{dn}) we have $\Delta_2(x)={\Delta^-}(x)^2-{\Delta^+}(x)^2$, and
from eq. (\ref{dfourier}) follows ${\Delta^-}(-x)=-{\Delta^+}(x)$, therefore we obtain
for the Fourier transform of $\Delta_2(x)$
\begin{equation}
{\hat \Delta}_2 (k)=\frac{1}{8 \pi} \mbox{sgn}(k^0) \Theta(k^2).
\end{equation}
It is obvious that ${\hat \Delta}_2(k)$ is related to the real part of eq. (\ref{gamma2}).

\section{Point splitting method in momentum space}
Instead of introducing a mass term as regulator of the distribution ${\hat \Delta}_2(k)$,
we present here another strategy, namely a point splitting
procedure in momentum space, which also leads to the desired result.
As was shown in \cite{eg,YMII},
the massless causal distribution ${\hat \Delta}_2(k)$ can be split in analogy to
eq. (\ref{disprel}) by computing a retarded part ${\hat \Delta}^q_{2,ret}(k)$
in momentum space of $\Delta_2(x) e^{iqx}$ in configuration
space via the {\em{subtracted}} dispersion relation \cite{ASchw}
\begin{equation}
{\hat \Delta}^q_{2,ret} (k) = \frac{i}{2 \pi} \int dt \frac{{\hat \Delta}_2(tk+q)}
{(t-i0)(1-t+i0)}
\end{equation}
for $k-q \! \in \! V^+$, $q^2 < 0$ and performing a limit
\begin{equation}
{\hat \Delta}_2^{ret} (k)=\lim \limits_{q \rightarrow 0} [{\hat \Delta}_{2,ret}^q (k)
+ c_q ] \, , \label{qlimit}
\end{equation}
where $c_q$ is a diverging $q$-dependent constant (i.e., $c_q$ is independent of $k$)
which can be chosen in such
a way that the limit in eq. (\ref{qlimit}) exists.
Note that ${\hat \Delta}_2(k+q)$ is the Fourier transform of
$\Delta_2(x) e^{iqx}$.
Therefore, one has to compute ($k \! \in \! V^+$)
\begin{equation}
{\hat \Delta}_2^{ret} (k)=\lim \limits_{q \rightarrow 0} \Biggl\{
\frac{i}{16 \pi^2} \int dt \, \frac{\mbox{sgn}(tk^0+q^0)
\Theta((tk+q)^2)}{(t-i0)(1-t+i0)} + c_q \Biggr\} \, . \label{split}
\end{equation}
The roots of $(tk+q)^2=t^2k^2+2t(kq)+q^2=0$ are
\begin{equation}
t_{1,2}=\frac{1}{k^2} (-kq \pm \sqrt{N})
\end{equation}
with $N=(kq)^2-k^2q^2 > (kq)^2$ and $t_1>0$, $t_2<0$.
The integral in (\ref{split}) becomes
\begin{displaymath}
\Biggl\{\frac{-i}{16 \pi^2} \int \limits_{-\infty}^{t_2} dt
+ \frac{i}{16 \pi^2} \int \limits_{t_1}^{\infty} dt \Biggr\}
\Bigl\{ \frac{1}{t} - P \frac{1}{t-1} - i \pi \delta(t-1) \Bigr\}
\end{displaymath}
\begin{equation}
=\frac{i}{16 \pi^2} (-\log |t_2| + \log |t_2-1| - \log |t _1 |
+\log|t_1-1| - i \pi).
\end{equation}
Since $t_{1,2} \rightarrow 0$ for $q \rightarrow 0$ the logarithms $\log |t_{1,2}-1|$
vanish in this limit and the integral reduces to
\begin{equation}
-\frac{i}{16 \pi^2} ( \log |t_1 t_2| + i \pi)= \frac{i}{16 \pi^2} \Bigl(
\log \frac{k^2}{|q^2|} - i \pi \Bigr). \label{finalsplit}
\end{equation}
Adding $c_q=\frac{i}{16 \pi^2} \log \frac{|q^2|}{\mu_{\mathcal{R}}^2}$
to eq. (\ref{finalsplit}) leads after analytic
continuation to the (covariant) splitting solution
\begin{equation}
{\hat \Delta}_2^{ret} (k)=\frac{i}{16 \pi^2} \log
\Biggl (-\frac{k^2+i k^0 0}{\mu_{\mathcal{R}}^2} \Biggr) \, ,
\end{equation}
in accordance with the well-known result for $\Gamma_2^{\mathcal{R}}$ given by
eq. (\ref{gamma2}).

\section{Results for higher loops}
We proceed with the Fourier transform of $\Delta^-(x)^3=:\Delta_3^- (x)$
\begin{displaymath}
{\hat \Delta}_3^-(k)=\frac{i}{(2 \pi)^3} \int d^4 q \, \Theta(-q^0) \delta(q^2)
{\hat \Delta}_2^-(k-q)
\end{displaymath}
\begin{equation}
=-\frac{i}{4 (2 \pi)^4} \int d^4 q \, \Theta(-q^0) \delta(q^2)
\Theta(q^0-k^0) \Theta((k-q)^2) \label{anchor}
\end{equation}
For $k=(k^0 \! < \! 0,\vec{0})$ we obtain
\begin{displaymath}
{\hat \Delta}_3^-(k)=-\frac{i}{4 (2 \pi)^4} \int \frac{d^3 q}{2 E} \,
\Theta(-E-k^0) \Theta((k^0)^2 +2 k^0 E)
\end{displaymath}
\begin{equation}
-\frac{i}{4(2 \pi)^3} \int \limits_{0}^{-k^0/2} \, d |\vec{q} \, | |\vec{q} \, |=
-\frac{i}{32(2 \pi)^3} k^2 \, , \label{recursion}
\end{equation}
hence
\begin{equation}
{\hat \Delta}_3^-(k)=-\frac{i}{32(2 \pi)^3} k^2 \Theta(-k^0) \Theta(k^2).
\end{equation}
Inserting a factor $(k-q)^2 = (k^0)^2+2 k^0 E$ in eq. (\ref{recursion}),
corresponding to a replacement ${\hat \Delta}_{n}^-(k) \rightarrow
{\hat \Delta}_{n+1}^-(k)$ in eq. (\ref{anchor}),
leads after a short inductive calculation to $(n \ge 2)$
\begin{equation}
{\hat \Delta}_n^-(k)=\frac{i^n (k^2)^{n-2}}{4^{n-1} (n-1)! (n-2)! (2 \pi)^{2n-3}}
\Theta(-k^0) \Theta(k^2)
\end{equation}
or
\begin{equation}
{\hat \Delta}_n(k)=\frac{(k^2)^{n-2}}{4^{n-1} (n-1)! (n-2)! (2 \pi)^{2n-3}}
\mbox{sgn}(k^0) \Theta(k^2).
\end{equation}
Keeping in mind that $(k^2)^{n-2}$ in Fourier space corresponds to
the local operator $(-\Box)^{n-2}$ in configuration space,
a retarded part of $\Delta_n$ can be obtained directly from the retarded
solution $\Delta_2^{ret}$ calculated above for $\Delta_2$ in momentum space,
since $(k^2)^{n-2}
{\hat \Delta}_2^{ret} (k)$ is a retarded part of $(k^2)^{n-2} {\hat \Delta}_2(k)$.
The actual calculation is straightforward and leads
after analytic continuation of ${\hat \Delta}_2^{ret} (k)$
to the desired result
\begin{equation}
\Gamma_n^{\mathcal{R}}(k) = -\frac{(-i)^{n+1} (k^2)^{n-2}}
{4^{n-1} (2 \pi)^{2n-2} (n-1)! (n-2)!} \log
\Biggl (-\frac{k^2+i0}{\mu_{\mathcal{R}}^2} \Biggr) + 
\sum \limits_{i=0}^{n-2} c_i^{\mathcal{R}}
(k^2)^i
\end{equation}
for $n \ge 2$.
\begin{acknowledgements}
The author wishes to thank Cyrill von Arx for generating the figure.
This work was supported by the Swiss National Science Foundation.
\end{acknowledgements}


\begin{thebibliography}{99}

\bibitem{Pauli}
Pauli, W., Villars, F.:
On the invariant regularization in relativistic quantum theory. 
Rev. Mod. Phys. {\bf{21}}, 434 (1949).

\bibitem{Dim}
't Hooft, G., Veltman, B.:
Regularization and renormalization of gauge fields.
Nucl. Phys. {\bf {B44}}, 189 (1972).

\bibitem{Schwinger}
Schwinger, J.:
On gauge invariance and vacuum polarization.
Phys. Rev. {\bf 82}, 664 (1951).

\bibitem{GScharf} Scharf, G.: {\it Finite Quantum Electrodynamics}.
2nd ed., Springer Verlag, New York, 1995.

\bibitem{eg}
Epstein H., Glaser V.: 
The role of locality in perturbation theory.
Annales Poincar\'{e} Phys. Theor. {\bf{A19}}, 211 (1973).

\bibitem{Aste} Aste, A.:
The two loop master diagram in the causal approach.
Annals Phys. {\bf 257}, 158 (1997).

\bibitem{Cheng}
Cheng, T.-P., Li, L.-F.: {\it Gauge Theory of Elementary Particle
Physics}. Oxford University Press, New York, 1984.

\bibitem{Kallen} K\"all\'en, G., Sabry, A.:
Fourth order vacuum polarization.
Dan. Mat. Fys. Medd. {\bf 29}, no. 17, 1. (1955).

\bibitem{TkaChet} Tkachov, F.V.:
A theorem on analytical calculability of four loop renormalization group functions.
Phys. Lett. {\bf B100}, 65 (1981);
Chetyrkin, K.G., Tkachov, F.V.:
Integration by parts: The algorithm to calculate beta functions in 4 loops.
Nucl. Phys. {\bf B192}, 159 (1981).

\bibitem{Bauberger}
Bauberger, S., Berends, F.A., B\"ohm, M., Buza, M.:
Analytical and numerical methods for massive two loop selfenergy diagrams.
Nucl. Phys. {\bf{B434}}, 383 (1995).

\bibitem{Berends} Berends, F.A., Davydychev, A.I., Ussyukina, N.I.:
Threshold and pseudothreshold values of the sunset diagram.
Phys. Lett. {\bf B426}, 95 (1998).

\bibitem{Fleischer} Fleischer, J., Kotikov, A.V., Veretin, O.L.:
Analytic two loop results for selfenergy type and vertex type
diagrams with one nonzero mass.
Nucl. Phys. {\bf B547}, 343 (1999).

\bibitem{PostTausk} Post, P., Tausk, J.B.:
The sunset diagram in SU(3) chiral perturbation theory.
Mod. Phys. Lett. {\bf A11}, 2115 (1996).

\bibitem{DavySmi} Davydychev, A.I., Smirnov, V.A.:
Threshold expansion of the sunset diagram.
Nucl. Phys. {\bf B554}, 391 (1999).

\bibitem{Caffo2} Caffo, M., Czy{\.z}, H., Remiddi, E.:
The threshold expansion of the two loop sunrise selfmass master amplitudes.
Nucl. Phys. {\bf B611}, 503 (2001).

\bibitem{Passarino} Passarino, G.:
An approach toward the numerical evaluation of multiloop Feynman diagrams.
Nucl. Phys. {\bf B619}, 257 (2001).

\bibitem{AsteSunrise} Aste, A., Trautmann, D.:
Finite calculation of divergent selfenergy diagrams.
Can. J. Phys. {\bf{81}}, 1433 (2003).

\bibitem{Laporta} Laporta, S., Remiddi, E.:
The analytic value of a 4-loop sunrise graph in a particular
kinematical configuration.
Nucl. Phys. {\bf B704}, 349 (2005).

\bibitem{Laporta2} Laporta, S.:
High-precision calculation of multiloop Feynman integrals by difference equations.
Int. J. Mod. Phys. {\bf A15}, 5087 (2000).

\bibitem{Groote} Groote, S., Pivovarov, A.A.:
Threshold expansion of Feynman diagrams within a configuration space technique.
Nucl. Phys. {\bf B580}, 459 (2000).

\bibitem{GKP} Groote, S., K\"orner, J.G., Pivovarov, A.A:
On the evaluation of a certain class of Feynman diagrams in x-space:
Sunrise-type topologies at any loop order. hep-ph/0506286.

\bibitem{PCT}
Streater, R.F., Wightman, A.S.: {\it PCT, Spin, Statistics and All
That}. Benjamin-Cummings Publishing Company, 1964.

\bibitem{Constantinescu}
Constantinescu, F.:
{\it Distributions and Their Applications in Physics}. Pergamon Press, 1980.
\bibitem{YMII}
D\"utsch, M., Hurth, T., Krahe, F., Scharf, G.:
Causal construction of Yang-Mills theories: II.
Nuovo Cim. {\bf{A107}}, 375 (1994).

\bibitem{ASchw}
Aste, A., G. Scharf, G., Walther, U.:
Power counting degree versus singular order in the Schwinger model.
Nuovo Cim. {\bf A111}, 323 (1998).

\end{thebibliography}
\end{document}